\def\slash#1{\setbox0=\hbox{$#1$}#1\hskip-\wd0\dimen0=5pt\advance
\dimen0 by-\ht0\advance\dimen0 by\dp0\lower0.5\dimen0\hbox
to\wd0{\hss\sl/\/\hss}}

\documentclass{JHEP}
\usepackage{epsfig}
\newcommand{\be}{\begin{equation}}
\newcommand{\ee}{\end{equation}}

\title{Couplings of pions to higher positive parity heavy mesons}

\author{A. Deandrea\\
Institut f\"ur Theoretische Physik, Universit\"at Heidelberg,\\
Philosophenweg 19, D-69120 Heidelberg, Deutschland}
\author{R. Gatto\\
D\'epartement de Physique Th\'eorique, Universit\'e de Gen\`eve,\\
24 quai E.-Ansermet, CH-1211 Gen\`eve 4, Suisse}
\author{G. Nardulli and A. D. Polosa\\
Dipartimento di Fisica, Universit\`a di Bari and INFN Bari,\\
via Amendola 173, I-70126 Bari, Italia}

\abstract{We evaluate the couplings of pions in the transitions  of
positive parity heavy mesons, $(1^+,2^+)$ and $(0^+,1^+)$, to negative 
parity ones $(0^-,1^-)$ using a technique which is not limited to the 
soft-pion limit. This is made through a constituent quark-meson model 
(the CQM model) where the amplitudes are obtained by quark loops with 
mesons on the external lines. The results are in good agreement with 
experimental data.}


\preprint{BARI-TH/98-327\\
HD-TVP-99-1\\
UGVA DPT 1999 - 01/1025}

\begin{document}

\section{Introduction}

Positive parity mesons comprising one heavy quark are the object of a
wide experimental search \cite{exp} and this demands a deeper theoretical
understanding of processes involving these states. A theoretical description
within QCD can be attempted at present only by
numerical simulations or by the use of QCD sum rules. However
much of what is known about QCD can also be described by using an effective
Lagrangian, encoding the symmetries and dynamical properties of the
theory with a limited number of parameters. This is possible
because the use of expansion parameters, such as the energy or the heavy
quark mass, makes sense in some limited kinematical regions, 
whereas in principle
the effective Lagrangian description would require an infinite number
of operators, and correspondingly an infinite number of free parameters.
The effective Lagrangian approach is usually simple enough to allow for
analytical calculations, thus giving the opportunity to see in detail
the interplay between symmetries and dynamical information introduced
into the model.

We are interested here in a constituent quark-meson model (the CQM model
\cite{gatto,IIgatto}) in the
intermediate range of energies between the confinement scale and the 1 GeV
region.  In this model both quark and meson effective fields are
present.
In the following we extend the analysis described in \cite{gatto,IIgatto}
to the study of strong decays $D_2^* \to D\pi$, $D_1^*\to D\pi$.
Here the states $D_2^*$, $D_1^*$ are the observed 2460 and 2420 charmed
positive parity mesons. We also
generalise the computation of strong decays $D_1^{*\prime} \to D \pi$,
$D_0 \to D^* \pi$ made in~\cite{gatto} avoiding
the soft-pion-approximation used there (the $D_1^{*\prime}$, $D_0$ states
have $J^P= 1^+$, $0^+$). 

Our strategy is to consider a quark-meson Lagrangian where transition
amplitudes are represented by diagrams with heavy mesons attached to loops
containing heavy and light constituent quarks. The effective meson-quark
vertices emerge as a result of path integral bosonization of an underlying
4-quark Nambu-Jona-Lasinio  (NJL) interaction.
Our results are encouraging especially in consideration of the technical
simplicity that the method offers in comparison to other available
approaches, such
as for instance in~\cite{aliev}.

\section{The model}

A short presentation of the CQM model is in order; for a full account
see \cite{gatto} and references therein. The model is an effective field
theory
containing a quark-meson Lagrangian incorporating heavy-quark
symmetries~\cite{neurev}. For our purposes this Lagrangian can be
divided into  two parts:
\begin{equation}
{\cal L} ={\cal L}_{\ell \ell} + {\cal L}_{h \ell}.
\label{lagra}
\end{equation}
The first term involves only the light quark degrees of freedom, $\chi$,
and the pseudoscalar
light mass octet. At the lowest order in a derivative expansion one has:
\begin{equation}
{\cal L}_{\ell \ell}={\bar \chi} (i  D^\mu \gamma_\mu +
{\cal A}^\mu \gamma_\mu \gamma_5) \chi - m {\bar \chi}\chi
+ {f_{\pi}^2\over 8} \partial_{\mu} \Sigma^{\dagger} \partial^{\mu}
\Sigma \; .
\end{equation}
Apart from the mass term, ${\cal L}_{\ell \ell}$ is chiral invariant.
Here $D_\mu = \partial_\mu-i {\cal V}_\mu$,
$\xi=\exp(i\pi/f_\pi )$, $\Sigma=\xi^2$, $f_\pi=130$ MeV.
$\pi$ is the $3 \times 3$ matrix representing the flavour $SU(3)$ octet of
pseudoscalar mesons and the explicit expressions for ${\cal V}^{\mu}$ and
${\cal A}^{\mu}$ are:
\begin{eqnarray}
{\cal V}^\mu &=& {1\over 2} (\xi^\dagger \partial^\mu \xi +\xi \partial^\mu
\xi^\dagger) \nonumber \\
{\cal A}^\mu &=& {i\over 2} (\xi^\dagger \partial^\mu \xi -\xi \partial^\mu
\xi^\dagger)~. \label{av}
\end{eqnarray}
As one can see by expanding the exponential of the pseudoscalar mesons
$\xi$, ${\cal V}^{\mu}$ generates couplings of an even number of
pseudoscalar mesons to the ${\bar \chi} \chi$ pair, while ${\cal A}^{\mu}$
gives an odd number of $\pi$ fields.

The effective Lagrangian containing heavy and light quarks together with
meson fields is:
\begin{eqnarray}
{\cal L}_{h \ell}&=&{\bar Q}_v i v\cdot \partial Q_v
-\left[ {\bar \chi}({\bar H}+{\bar S}+ i{\bar T}_\mu
{D^\mu \over {\Lambda_\chi}})Q_v +h.c.\right]\nonumber \\
&+&\frac{1}{2 G_3} {\mathrm {Tr}}[({\bar H}+{\bar S})(H-S)]
+\frac{1}{2 G_4}
{\mathrm {Tr}} [{\bar T}_\mu T^\mu ]~~.
\label{qh1}
\end{eqnarray}
Here $Q_v$ is the effective heavy quark field of Heavy Quark
Effective Theory (HQET)~\cite{neurev}, $\Lambda_{\chi}$ is the
momentum scale characterising the convergence of the derivative expansion,
usually taken as the chiral symmetry breaking scale $\Lambda_{\chi}\simeq
1$~GeV, and $H$, $S$, $T$ fields are the effective meson fields
discussed in~\cite{falk} which,
in the explicit matrix representation, have the form:
\begin{eqnarray}
H &=& \frac{1+\slash v}{2}\; [P_{\mu}^*\gamma^\mu - P \gamma_5 ]\\
S &=& {{1+\slash v}\over 2}[P_{1\mu}^{*\prime} \gamma^\mu\gamma_5-P_0]\\
T^\mu &=& {\frac {1+\slash v}{2}}\left[P_2^{* \mu\nu}\gamma_\nu-\sqrt{\frac 3
2}
P^*_{1\nu}\gamma_5\left(g^{\mu\nu}-\frac 1 3
\gamma^\nu(\gamma^\mu-v^\mu)\right)
\right] \; .
\end{eqnarray}
Here $P,P^*_{\mu}$ etc. are annihilation operators normalised in the
following way:
\begin{eqnarray}
\langle 0|P| Q{\bar q} (0^-)\rangle & =&\sqrt{M_H}\\
\langle 0|{P^*}^\mu| Q{\bar q} (1^-)\rangle & = & \epsilon^{\mu}\sqrt{M_H}\\
\langle 0|{P_2^*}^{\mu \nu}| Q{\bar q} (2^+)\rangle & = & \epsilon^{\mu\nu}
\sqrt{M_T}
\end{eqnarray}
$H, S$ and $ T^{\mu}$ describe respectively the $(0^-,1^-)$ doublet, the
$(0^+,1^+)$ doublet $D_0,D_1^{*\prime}$ and the $(1^+,2^+)$ doublet
$D_1^*,D_2^*$ that are predicted by HQET.
The part in the Lagrangian containing the $T$ field is not directly  read from
the underlying NJL contact interaction, but it constitutes its expected
extension.
The coupling constants $G_3$, $G_4$ and the field renormalization
constants $Z_H, Z_S, Z_T$ are discussed in \cite{gatto} (for numerical
values see below). Another parameter of the model is $\Delta_H$, i.e.
the difference between the $H$ meson doublet mass and the mass of the
heavy quark involved.
This parameter is not predicted by the model but, once $\Delta_H$ is fixed,
the model gives $\Delta_S$ (as a result of the dynamical information
introduced into the Lagrangian by the underlying NJL contact interaction),
while $\Delta_T$ is still a free parameter that can be estimated using
experimental information. The following table was computed in \cite{gatto}:
\TABULAR[htb]{|c|c|c|}
{\hline
$\Delta_H$ & $\Delta_S$ & $\Delta_T$ \\
\hline
$0.3$ & $0.545$ & $0.635$ \\
\hline
$0.4$ & $0.590$ & $0.735$ \\
\hline
$0.5$ & $0.641$ & $0.835$ \\
\hline}
{$\Delta$ values (in GeV)}
We allow for a variation of $\Delta_H=0.4\pm 0.1$~GeV, since only this
range of
values is allowed by the results on semileptonic weak decays (for a
discussion see \cite{gatto}).

The last step necessary to the definition of the model is the choice of a
regularization procedure.
We use the Schwinger proper time regularization method assuming as UV and IR
cut-off  $\Lambda\simeq 1.25$~GeV and $\mu\simeq 0.3$~GeV respectively.
The former ensures the suppression of large momentum flows through light quark
lines in the loops, since the heavy quark carries most of the energy and
momentum in the system; the latter, being of order of $\Lambda_{\rm QCD}$,
drops the unknown confinement part of the
quark interaction (see for example \cite{ebert}).
The regularized form of the heavy quark propagator is:

\begin{equation}
\int d^4 k_E  \frac{1}{k_E^2+m^2} \to \int d^4 k_E \int_{1/
\Lambda^2}^{1/\mu^2} ds\; e^{-s (k_E^2+m^2)}~~, \label{cutoff}
\end{equation}
where $m$ is the constituent light quark mass, which we take as $m=0.3$ GeV
for $u$ and $d$
flavours. An alternative way of cutting off high momenta is proposed
in~\cite{holdom}.

\section{Strong couplings}

As shown in~\cite{falk}, transitions $T\to H\pi$ are described by an
interaction Lagrangian that in HQET has the the following form:
\begin{equation}
{\cal L}=\frac{h_1}{\Lambda_{\chi}} {\rm Tr}[{\bar H}T^{\mu}(iD_{\mu}
\slash{{\cal  A}})\gamma_5] + \frac{h_2}{\Lambda_{\chi}} {\rm Tr}[{\bar
H}T^{\mu}
(i \slash{D}  {\cal A}) \gamma_5] + h.c. \; ,
\label{lagt}
\end{equation}
while for the $S\to H\pi$ transitions we have:
\begin{equation}
{\cal L}=ig {\mathrm {Tr}}(\overline{H} H \gamma^{\mu}\gamma_5{\cal A}_{\mu})
\; +\; \left[ih\; {\mathrm {Tr}}(\overline{H} S \gamma^{\mu}\gamma_5
{\cal A}_{\mu})  \; + \; {\mathrm {h.c.}}\right]
\label{lg}
\end{equation}
The general amplitudes, for the processes we are interested in, are defined
as follows:
\begin{eqnarray}
\langle D^{+}(p^{\prime}) \pi^{-}(q)| D_2^{*0}(p,\eta) \rangle &=&
ig_1 \eta^{\mu\nu} q_\mu q_\nu
\label{mat1}\\
\langle D^{*+}(p^{\prime},\epsilon) \pi^{-}(q)| D_2^{*0}(p,\eta) \rangle &=&
ig_2 \eta^{\mu\nu}
q_\mu \epsilon_{\lambda\sigma\nu\tau}\epsilon^{*\lambda}\frac{p^{\sigma}}{m_T}
q^{\tau}
\label{mat2}\\
\langle D^{*+}(p^{\prime},\epsilon) \pi^{-}(q)| D_1^{*0}(p,\eta) \rangle &=&
ig_3\eta_{\nu}\epsilon^{*}_{\sigma} [3q^{\nu}q^{\sigma}+g^{\nu\sigma}
(q^2-\frac{(p\cdot q)^2}{m_T})]
\label{mat3}\\
\langle D^{0}(p^{\prime}) \pi^{+}(q)| D_0^+(p) \rangle &=&
ig_4\frac{m_S^2-m_H^2}{m_S}
\label{mat4}\\
\langle D^{0}(p^{\prime}) \pi^{+}(q)| D^{*+}(p,\epsilon) \rangle &=&
ig_5q^{\mu}\epsilon_{\mu} \; ,
\label{mat5}
\end{eqnarray}
where we are using the HQET definitions of the symbols $H,S,T$ for labelling
mesons masses. The connection with $h_1,h_2,g,h$ is:
\begin{eqnarray}
g_1& = & g_2 = 2\sqrt{m_H m_T} \frac{h^{\prime}}{\Lambda_{\chi} f_{\pi}}\\
g_3 & = & \sqrt{\frac{2 m_H m_T}{3}}\frac{h^{\prime}}{\Lambda_{\chi}
f_{\pi}}\\
g_4 & = & \sqrt{m_H m_S}\frac{h}{ f_{\pi}}\\
g_5 & = & \frac{2 m_H}{f_\pi} g~~,
\end{eqnarray}
where $h^{\prime}= h_1+h_2$.

Our method consists in a loop calculation of a diagram such as in Fig. 1
involving
a current and two meson states on the external legs. Let us consider the
calculation of
$T\to H \pi$ , i.e. the determination of $h^\prime$.

According to the Feynman rules derived from Eq.~(\ref{lagra}) and
Eq.~(\ref{lagt}), the loop shown in figure may be written as:
\begin{equation}
-\frac{i}{2}\sqrt{Z_H Z_T m_H m_T} q^\mu \eta^{\sigma\nu}
\frac{iN_c}{16\pi^4} \int^{\mathrm {reg}}d^4k
\frac{{\rm Tr}[(\slash k + m)\gamma_\mu \gamma_5 (\slash k^{\prime} + m)
\gamma_5 (1+\slash v) \gamma_\nu k_\sigma]
}{(k^2-m^2)[(k+q)^2-m^2](v\cdot k + \Delta + i\epsilon)}
\label{mostro}
\end{equation}
The computation of the previous expression can be decomposed in simpler
expressions
involving integrals such as as $Z$, $Z^{\sigma}$, $Z^{\sigma\nu}$,
and $Z^{\sigma\nu\lambda}$ (reported in the appendix).
The numerators of these integrals are the Lorentz structures $1$,
$k^{\sigma}$, $k^{\sigma}k^{\nu}$, $k^{\sigma}k^{\nu}k^{\lambda}$
respectively.

\EPSFIGURE[h]{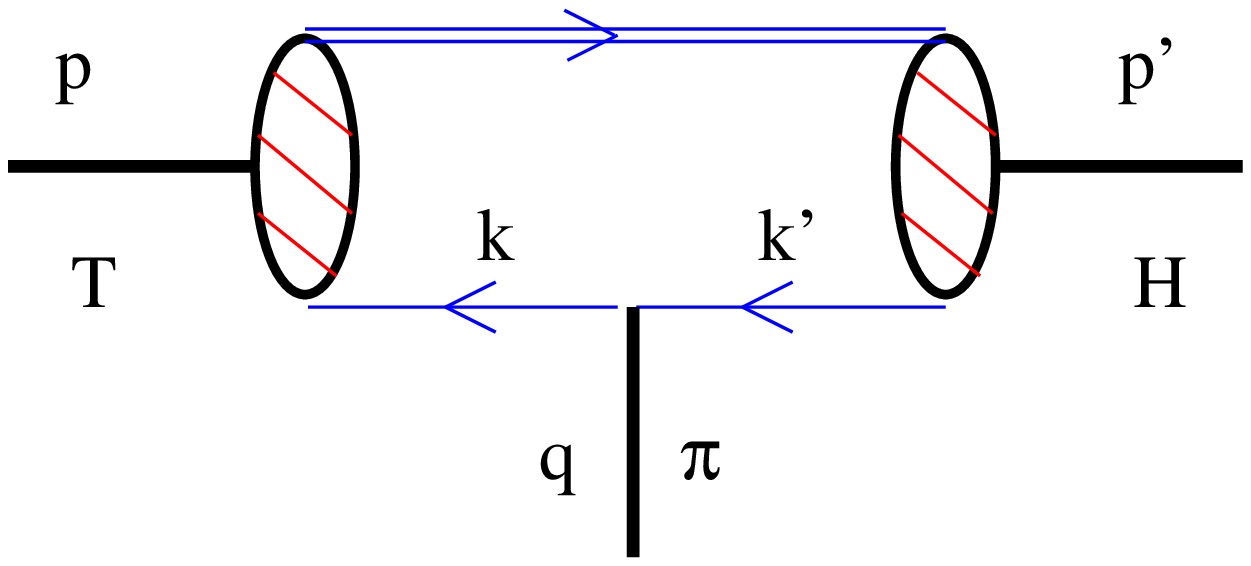,width=7cm}
{Diagram for the loop calculation of the $T\to H \pi$
coupling $h^{\prime}$ in the constituent quark-meson model.}

Once Eq.~(\ref{mostro}) is calculated, the coupling $h^\prime$ may
be obtained by matching the previous calculation in the quark-meson
loop model Eq.~(\ref{qh1}) with the expression of the effective Lagrangian,
where the only degrees of freedom are heavy and light me\-sons
Eq.~(\ref{lagt}) and $h' = h_1+h_2$. The same strategy is followed for
the evaluation of $h$ and $g$ in Eq.~(\ref{lg}).
For the $T\to H \pi$ coupling we find:
\begin{eqnarray}
h^\prime &=& \sqrt{Z_T Z_H} \Big\{ \frac{m^2}{q_\pi} \Big[I_2+\Delta_T
Z(\Delta_T )+ \frac{1}{2 q_\pi} (I_3(\Delta_H)-I_3(\Delta_T))\Big]
\nonumber \\
&+& P(R_i(\Delta_T),S_i(\Delta_T),q_\pi)\Big\} ~, \label{acc1}
\end{eqnarray}
where the polynomial $P(R_i,S_i,q_\pi)$, expressed as a sum of powers of
$q_\pi$ multiplied by the linear combinations $R_i$ and $S_i$ of $I_i$ and $Z$
integrals, is given in the appendix, together with the expressions
for $I_i$ and $Z$. The parameter $q_\pi$ is $q_\pi=v\cdot q$ where
$v$ is the velocity of the incoming meson and $q^\mu$ is the pion 4-momentum.
Consistently with chiral invariance we neglect the pion mass and we write
$q^\mu=(q_\pi,0,0,q_\pi)$ in the decaying particle rest frame. Also,
consistently
with the effective Lagrangian (\ref{lagt}),  containing the minimum number
of couplings, in the evaluation of Eq. (\ref{acc1}) we neglected $O(q_\pi^3)$
terms. Two powers of $q_\pi$ have to be kept, since they appear in the
D-wave matrix element (\ref{mat1})-(\ref{mat3}).

Varying $\Delta_H$ in the same range as in Table I,
we can evaluate the  variation of the field renormalization
constants $Z_H, Z_S, Z_T$ defining the renormalized fields
${\hat H}=H/\sqrt{Z_H}$, ${\hat S}=S/\sqrt{Z_S}$ and ${\hat T}=T/\sqrt{Z_T}$.
We only report values of $Z_H, Z_S, Z_T$ for $\Delta_H=0.4$ which are
respectively $2.36, 1.14, 1.07~{\rm GeV}^{-1}$ (see \cite{gatto} for a
complete table).

For the strong coupling  $S\to H \pi$ we find:
\begin{equation}
h =\sqrt{Z_S Z_H}\left[R_1(\Delta_S)- R_2(\Delta_S)-\frac{R_4(\Delta_S)}
{q_\pi}+ m^2 Z(\Delta_S)\right].
\label{acca}
\end{equation}
In this calculation we have considered terms only up to the order $O(q_\pi^1)$,
to be consistent with the energy expansion for the chiral
fields, as explained before; this result goes beyond the soft pion limit
approximation (SPL) which is often used in these calculations (see the
discussion below). Numerically we find:
\begin{equation}
h^\prime  =0.65^{+0.45}_{-0.3}\; ,
\end{equation}
where the error is induced by the variation of $\Delta_H$ in the range
$0.3-0.5$~GeV. For $h$ we obtain:
\begin{equation}
h=-0.76\pm 0.13 \; .
\end{equation}
Here we use the mass determination of the $S-$multiplet made in \cite{gatto}
i.e. $2.165\pm 0.05$~GeV. The error in this determination does not
contribute significantly to $h$. Note that recent preliminary experimental 
results by CLEO \cite{tracleo} give a mass of $2.461$ GeV for the broad
$1^+$ state attributed to the $S$ multiplet. This difference may be understood
as a consequence of a ${\cal O}(1/m_c)$ correction to the mass formula. 

Using the preliminary CLEO results for the $1^+$ mass in the $S$ multiplet
(it affects the pion momentum in (\ref{acca})), we get:
\begin{equation}
h=-0.56\pm 0.11
\label{newh}
\end{equation}
which is consistent with the one obtained by QCD sum rules~\cite{col}:
\begin{equation}
h=-0.52\pm 0.17\; .
\end{equation}
The approach used in this paper to calculate the strong coupling constants
outside the SPL is an improvement with respect to \cite{gatto}, where we
found $h=-0.85\pm 0.02$ using $q_\pi \to 0$ from the very beginning.
For $g$, the soft pion limit is instead a very good approximation since the
decay $D^* \to D \pi$ has a very limited phase space. This is confirmed by
the evaluation of $g$ made in \cite{gatto}:
\begin{equation}
g=0.46 \pm 0.04\; .
\end{equation}
This value agrees with QCD sum rules calculations \cite{rep} and with
a result from a relativistic constituent quark model \cite{defazio},
giving $g=0.44\pm 0.16$ and $g=0.40$ respectively.

We conclude by calculating the widths of the $T$ and $S$ states 
to $H$ plus pion, where the heavy quark is a c quark, since these
states are already observed. These two body decay processes, from the
chiral  heavy meson Lagrangian, were already discussed in \cite{falk},
to which we refer for more details. Using updated experimental
data \cite{pdg} for the masses in the phase space integrals, and our
determination of the coupling $h^\prime$, we obtain:
\begin{eqnarray}
\Gamma (D_2^{*0} \to D^+ \pi^-) &=& 4.59 \times 10^7
\frac{h^{\prime 2}}{\Lambda_\chi^2} MeV = 19.4 MeV\\
\Gamma (D_2^{*0} \to D^{*+} \pi^-) &=& 1.33 \times 10^7
\frac{h^{\prime 2}}{\Lambda_\chi^2} MeV = 5.6 MeV\\
\Gamma (D_1^{*0} \to D^{*+} \pi^-) &=& 1.47 \times 10^7
\frac{h^{\prime 2}}{\Lambda_\chi^2} MeV = 6.2 MeV
\label{width21}
\end{eqnarray}
in which the central value of $h^\prime = 0.65$ is used and
$f_\pi =130$ MeV, $\Lambda_\chi = 1000$ MeV. The full one pion
widths are a factor 1.5 larger than those in the previous formula because
of the neutral pion decay channel. As the total $D_2^{*0}$ width is
dominated by the one pion mode in the chiral heavy meson Lagrangian,
one can use the experimental result of $23 \pm 5$ MeV to extract an
experimental value for $h^\prime$ from the previous formula. One obtains a
value of 0.51, which is in good agreement with the value of 
$h^\prime$ predicted by our model.
From (\ref{width21}) one can also deduce the total one pion width
of $D_1^{*0}$, $\Gamma = 9.3$ MeV, which is only half the measured
total with $18.9^{+4.6}_{-3.5}$. However this state can mix with the
broader $1^+$ state in the $S$ multiplet, the $D_1^\prime$ or can be
affected by significant $1/m_c$ corrections \cite{falkme}.
Neglecting this mixing we can compute the width of the $D_1^\prime$
state by using the preliminary CLEO result for the mass \cite{tracleo} 
and the computed strong coupling constant $h$ given in (\ref{newh}).
We obtain 
\begin{equation}
\Gamma(D_1^\prime \to D^* \pi)=240\; {\mathrm MeV}~~, 
\end{equation}
which accounts probably for most of the total $D_1^\prime$ width due to the 
limited available phase space. This result can be compared with the 
preliminary CLEO data for the total width, $290^{+101}_{-79}\pm 26 \pm 36$
MeV.

\section{Conclusions}
In this paper we have investigated the phenomenological implications for
excited positive parity heavy mesons of the model
proposed in \cite{gatto}. The work is experimentally
relevant at present for the charm states, for which these transitions
are seen, and a comparison with theoretical calculations is indeed needed.
We have improved the technique for diagrams involving a pion vertex on the
light quark line, achieving a determination of the strong couplings of the
pion in the processes $S \to H\pi$ which turn out to be very close to the
QCD sum rule determinations \cite{col}.
Moreover, using a technique which goes beyond the soft-pion-limit
approximation, we have calculated the strong coupling of the pion
in the transition $T\to H \pi$, where such a  limit could not be reliably
applied. This coupling is then used to predict, for charmed heavy mesons, 
the decay widths of the $T$
doublet of states $2^+$ and $1^+$ to the $H$ states $1^-$ and $0^-$
plus one pion.

\acknowledgments

A.D. acknowledges the support of the EC-TMR (European Community
Training and Mobility of Researchers) Program on ``Hadronic Physics with
High Energy Electromagnetic Probes''. This work has also been carried
out in part under the EC program Human Capital and Mobility, contract UE
ERBCHRXCT940579 and OFES 950200. A.D. would like to thank Dr. Ziwei Lin
for pointing out a misprint in the $h^\prime$ value obtained from
the experimental widths in the last section of the paper.

\appendix

\section{Appendix}

In the paper we use several integrals that we list in this appendix.
\begin{eqnarray}
I_1&=&\frac{iN_c}{16\pi^4} \int^{\mathrm {reg}} \frac{d^4k}{(k^2 - m^2)}
={{N_c m^2}\over {16 \pi^2}} \Gamma(-1,{{{m^2}}
\over {{{\Lambda}^2}}},{{{m^2}}\over {{{\mu }^2}}})
\\
I_2&=&-\frac{iN_c}{16\pi^4} \int^{\mathrm {reg}} \frac{d^4k}{(k^2 - m^2)^2}
={{N_c }\over {16 \pi^2}} \Gamma(0,{{{m^2}}
\over {{{\Lambda}^2}}},{{{m^2}}\over {{{\mu }^2}}})
\\
I_3(\Delta) &=& - \frac{iN_c}{16\pi^4} \int^{\mathrm {reg}}
\frac{d^4k}{(k^2-m^2)(v\cdot k + \Delta + i\epsilon)}\nonumber \\
&=&{N_c \over {16\,{{\pi }^{{3/2}}}}}
\int_{1/{{\Lambda}^2}}^{1/{{\mu }^2}} {ds \over {s^{3/2}}}
\; e^{- s( {m^2} - {{\Delta }^2} ) }\;
\left( 1 + {\mathrm {erf}} (\Delta\sqrt{s}) \right)\\
I_4(\Delta)&=&\frac{iN_c}{16\pi^4}\int^{\mathrm {reg}}
\frac{d^4k}{(k^2-m^2)^2 (v\cdot k + \Delta + i\epsilon)} \nonumber\\
&=&\frac{N_c}{16\pi^{3/2}} \int_{1/\Lambda^2}^{1/\mu^2} \frac{ds}{s^{1/2}}
\; e^{-s(m^2-\Delta^2)} \; [1+{\mathrm {erf}}(\Delta\sqrt{s})]~.
\end{eqnarray}
where $\Gamma$ is the generalised incomplete gamma function and erf
is the error function.
We also define:
\begin{eqnarray}
Z(\Delta) &=&  \frac{iN_c}{16\pi^4} \int^{\mathrm {reg}}
\frac{d^4k}{(k^2-m^2)[(k+q)^2-m^2](v\cdot k + \Delta + i\epsilon)}\nonumber \\
&=&\frac{N_c}{16\pi^{3/2}} \int_{1/\Lambda^2}^{1/\mu^2} \frac{ds}{s^{1/2}}
e^{-s m^2} \int_{0}^{1}dx e^{s \Delta^2(x)} [1+{\mathrm {erf}}
(\Delta (x)\sqrt{s})]
\end{eqnarray}
Where $q^{\mu}=(q_\pi,0,0,q_\pi)$ is the pion 4-momentum and
$\Delta (x)= \Delta-xq_\pi$.
We can observe that the soft pion limit of the preceding expression, i.e.
the limit $q_\pi \to 0$, brings $Z(\Delta) \to I_4(\Delta)$ as it should be,
(cfr. I). We need  another auxiliary integral and some expressions which
are linear combination of the integrals listed. They are the following:
\begin{eqnarray}
\Theta &=& \frac{N_c}{16\pi^{2}}\int_{1/\Lambda^2}^{1/\mu^2}ds
\left( \frac{3-2 q_\pi^2 s}{6 s^2} \right)e^{-s m^2}  \\
R_1(\Delta_T) &=& m^2 Z(\Delta_T)-I_3(\Delta_H)\\
R_2(\Delta_T) &=& \Delta^2_T Z(\Delta_T)+\left(\frac{q_\pi}{2}+\Delta_T\right)I_2 \\
R_3(\Delta_T) &=& \frac{q_\pi}{2}(\Delta_T I_3(\Delta_T) -\Delta_H
I_3(\Delta_H))  \\
R_4(\Delta_T) &=& \frac{\Delta_T}{2}(I_3(\Delta_T)-I_3(\Delta_H)) \\
S_1(\Delta_T) &=& \Theta - \frac{\Delta_T h}{2} I_2 -\Delta^2_T I_2-\Delta^3_T
Z(\Delta_T)\\
S_2(\Delta_T) &=& I_1 + \Delta_T I_3(\Delta_H) - m^2 I_2 -m^2\Delta_T 
Z(\Delta_T) \\
S_3(\Delta_T) &=& q_\pi(I_1+\Delta_H I_3(\Delta_H))+\frac{m^2}{2}
(I_3(\Delta_H)-I_3(\Delta_T))\\
S_4(\Delta_T) &=& \frac{q_\pi}{2} I_1 + \frac{\Delta^2_T}{2}(I_3(\Delta_H) -
I_3(\Delta_T))\\
S_5(\Delta_T) &=& \frac{q_\pi \Delta_T}{2} (\Delta_H I_3(\Delta_H)-
I_3(\Delta_T))
\end{eqnarray}
where $q_\pi=v\cdot q$.\newline
We next define the following polynomial:
\begin{eqnarray}
P(R_i,S_i,q_\pi)&=&-\frac{1}{88 q_\pi^4}\left[8q_\pi^3(11 m R_1 + 4 S_1 - 6
S_2)+
2q_\pi^2(-176 m R_4
+14 S_1 + S_2 \right.\nonumber \\
&+& \left.8 S_3 + 48 S_4) +3q_\pi (88 m R_3 + S_3 - 16 S_4 - 32 S_5) 
+ 15 S_5 \right]
\end{eqnarray}
which is used in the equation for $h^\prime$.

\end{document}